\RequirePackage[hyphens]{url}
\documentclass{egpubl}
\usepackage{eurovis2020}
\usepackage[utf8]{inputenc}
\WsSubmission      % uncomment for submission to EG Workshop
\usepackage[T1]{fontenc}
\usepackage{dfadobe}  

\usepackage{cite}  % comment out for biblatex with backend=biber
\BibtexOrBiblatex
\electronicVersion
\PrintedOrElectronic
\ifpdf \usepackage[pdftex]{graphicx} \pdfcompresslevel=9
\else \usepackage[dvips]{graphicx} \fi

\usepackage{egweblnk}
\usepackage{enumitem}

\newcommand{\subscript}[2]{$#1 _ #2$}

\newenvironment{enumerate*}%
  {\begin{enumerate}[label=\subscript{G}{{\arabic*}}]%
    \setlength{\itemsep}{1.5pt}%
    \setlength{\parskip}{2pt}}%
  {\end{enumerate}}

\title[Tales from the Trenches]%
      {Tales from the Trenches: Developing sciview, a new 3D viewer for the ImageJ community}

\author[U. Günther \& K.I.S. Harrington]
{\parbox{\textwidth}{\centering Ulrik Günther$^{1,2,3}$\orcid{0000-0002-1179-8228} and
        Kyle I.S. Harrington$^{4}$\orcid{0000-0002-7237-1973}
        }
        \\
{\parbox{\textwidth}{\centering $^1$CASUS -- Center for Advanced Systems Understanding, G\"orlitz, Germany\\
         $^2$Max Planck Institute of Molecular Cell Biology and Genetics, Dresden, Germany\\
         $^3$Center for Systems Biology, Dresden, Germany\\
         $^4$Virtual Technology \& Design, University of Idaho, Moscow, Idaho, USA
       } 
}
}
\begin{document}

%\teaser{
% \includegraphics[width=\linewidth]{figures/teaser-drosophila.png}
% \centering
%  \caption{Volume rendering of a \emph{Drosophila} embryo from a 60 GiB dataset %using sciview.\TODO{Find/make better image?}}
%\label{fig:teaser}
%}

\maketitle
%-------------------------------------------------------------------------
\begin{abstract}
ImageJ/Fiji is a widely-used tool in the biomedical community for performing everyday image analysis tasks. However, its 3D viewer component
(aptly named \emph{3D Viewer}) has become dated and is no actively maintained. We set out to create an alternative tool that not only brings modern concepts and APIs from computer graphics to ImageJ, but is designed to be robust to long-term, open-source development. To achieve this we divided the visualization logic into two parts: the rendering framework, \emph{scenery}, and the 
user-facing application \emph{sciview}. In this paper we describe the development process and design decisions made, putting
an emphasis on sustainable development, community building, and software engineering best practises. We highlight the motivation for the Java Virtual Machine (JVM) as a target platform for visualisation applications. 
We conclude by discussing the remaining milestones and strategy for long-term sustainability.
%-------------------------------------------------------------------------
%  ACM CCS 1998
%  (see https://www.acm.org/publications/computing-classification-system/1998)
% \begin{classification} % according to https://www.acm.org/publications/computing-classification-system/1998
% \CCScat{Computer Graphics}{I.3.3}{Picture/Image Generation}{Line and curve generation}
% \end{classification}
%-------------------------------------------------------------------------
%  ACM CCS 2012
%The tool at \url{http://dl.acm.org/ccs.cfm} can be used to generate
% CCS codes.
%Example:
\begin{CCSXML}
<ccs2012>
<concept>
<concept_id>10011007.10011074.10011092</concept_id>
<concept_desc>Software and its engineering~Software development techniques</concept_desc>
<concept_significance>300</concept_significance>
</concept>
<concept>
<concept_id>10010147.10010371.10010372</concept_id>
<concept_desc>Computing methodologies~Rendering</concept_desc>
<concept_significance>300</concept_significance>
</concept>
<concept>
<concept_id>10010147.10010371.10010387</concept_id>
<concept_desc>Computing methodologies~Graphics systems and interfaces</concept_desc>
<concept_significance>300</concept_significance>
</concept>
</ccs2012>
\end{CCSXML}

\ccsdesc[300]{Software and its engineering~Software development techniques}
\ccsdesc[300]{Computing methodologies~Rendering}
\ccsdesc[300]{Computing methodologies~Graphics systems and interfaces}

\printccsdesc   
\end{abstract}
%-------------------------------------------------------------------------
\section{Introduction}

Scientific image processing and analysis is common throughout the scientific and engineering disciplines. While there are numerous software tools that support scientific image processing, one of leading open-source tools is ImageJ\cite{Schneider:2012nihi}. ImageJ is a Java-based tool that dates back to 1997\footnote{ImageJ is the Java-based successor to the NIH Image software package.}, and continues to be developed to this date. ImageJ was developed using Java to facilitate portability between systems. Additional features, such as a plugin system, have contributed to ImageJ's longevity. However, a number of limitations inherent to ImageJ's design were revealed in the 2010s, such as a fragile plugin ecosystem, and limited support for large images.

To alleviate these downsides of ImageJ, the Fiji (Fiji Is Just ImageJ) distribution of ImageJ \cite{schindelin2012fiji} was developed. Fiji introduced an update site mechanism for managing and distributing ImageJ plugins. Fiji also introduced support for a new backend library for image support, ImgLib2 \cite{Pietzsch:2012img}. The Fiji distribution revolutionized the ImageJ community by making numerous ImageJ plugins readily accessible within the tool itself. This is accomplished via \emph{update sites}, where the user can easily select plugin sources and manage plugin updates using a GUI, as opposed to the original approach which involved browsing to websites, downloading a precompiled JAR or compiled Java class, and installing it into ImageJ. Additionally, the introduction of the ImgLib2 library enabled support for the large scale image data that has now become popular, where original ImageJ data structures were limited to $2^{31}$ pixels, corresponding to the maximum integer value that can be used to index into an array of pixel data. Fiji also introduced the original 3D viewer\cite{Schmid:2010gm}.

%[cite imagej2 ecosystem paper]%https://www.ncbi.nlm.nih.gov/pmc/articles/PMC5428984/

ImageJ and the Fiji distribution of ImageJ continue to be popular tools for scientific image processing and analysis. This includes the use of ImageJ for 3D image analysis and visualization. However, recent advances in the visualization field have not yet made it into the Fiji ecosystem. To this end we designed and developed scenery and sciview.

\section{Motivations and Technical Aspects}

scenery itself was originally developed to prototype VR interactions with scientific instruments, such as high-speed volumetric microscopes. We found however, it is also a good basis for renewing the 3D visualisation infrastructure in ImageJ. Considering that biologists have the need to visualise larger and larger image data, generated by microscopes that can easily generate images of a rate of 1GB/s or higher \cite{Reynaud:2015dx}, with the total size often reaching into the terabytes, out-of-core volume rendering becomes a necessity. As scenery is agnostic of the frontend, we set out to develop sciview as the corresponding ImageJ plugin, which in turn depends on scenery.

Apart from out-of-core volume rendering, we thought the VR support already present in scenery might be of more general use for the biomedical community. While the original 3D Viewer only supported OpenGL 1.2, we also aimed for support of more up-to-date APIs -- a major reason for the abandonment of the original 3D Viewer was its strong dependency on Java3D, which has not been updated since Java 6 was released in 2006. In addition to supporting more up-to-date APIs, we observed that it would be beneficial to design a rendering framework that can utilize multiple APIs, and is not strongly tied to a single API.

The full goals for the development of scenery are \cite{Gunther:2019scenerya}:

\begin{enumerate*}
    \item \textbf{Virtual/Augmented Reality support}: The framework should make the use of VR/AR in an application possible with minimal effort. Distributed systems, such as CAVEs or Powerwalls, should also be supported.

    \item \textbf{Out-of-core volume rendering}: The framework should be able to handle datasets that do not fit into graphics memory and/or main memory, contain multiple channels, views, and timepoints. It should be possible to visualize multiple such datasets in a single scene.

    \item \textbf{User/Developer-friendly API}: The framework should have a simple API that makes only limited use of advanced features, such as generics, so the user/developer can quickly comprehend and customize it.

    \item \textbf{Cross-platform}: The framework should run on the major operating systems: Windows, Linux, and macOS.   
    %\item \textbf{Free and open-source software}: The framework needs to be available to all researchers for free, and all %of its source code needs to be available for inspection and modification.
    %Open-source software can boost reproducibility (reference? %https://www.nature.com/articles/nmeth.2082)
    %) and makes customisations easier.

    \item \textbf{JVM-native and embeddable}: The framework should run natively on the JVM, and be embeddable, such that it can be used in popular biomedical image analysis tools like Fiji \cite{schindelin2012fiji,Rueden:2017ij2}, Icy \cite{Chaumont:2012icy}, and KNIME \cite{Berthold2008}.
\end{enumerate*}

While scenery focuses on establishing a rendering framework based upon the above goals, sciview translates these goals into interfaces that are accessible from ImageJ. Two key efforts of sciview involve making data from ImageJ available to scenery for rendering (volumes, meshes, point clouds, etc.) and introducing new algorithms for both image and mesh processing. To this end, sciview primarily consumes ImageJ and ImgLib2 data structures, and generates/manipulates scenery data structures for visualisation. 

%The need for 3D visualization in bioimage analysis. 
% ImageJ's 3D viewer [cite bene's paper]

\begin{figure}
    \centering
    \includegraphics[width=\linewidth]{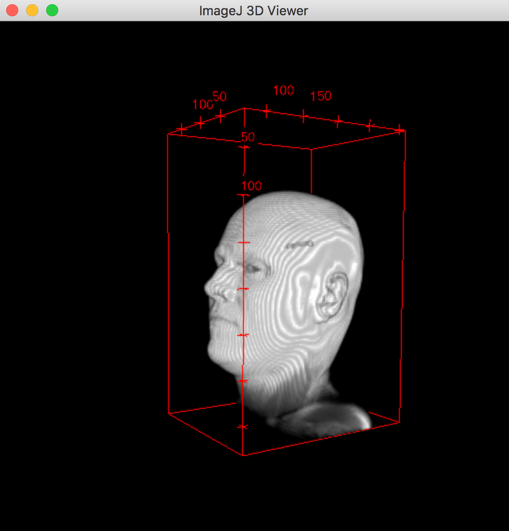}
    \caption{Example volume rendering from Fiji's original 3D viewer.}
    \label{fig:my_label}
\end{figure}

% Need for new features: 
% Vulkan
% VR
% out of core volume rendering: new innovations in microscopy has led to the development of new instruments that can easily acquire many GB per second and single image volumes on the order of TB. Visualizing the entire volume is often crucial for allowing the biologist to establish an intuition about the experimental sample, therefore techniques that enable out-of-core volume rendering are important.

\section{Approach}

With ImageJ being Java-based, the choice of target platform was already fixed, but the actual language was not: while the new viewer would need to run on the Java Virtual Machine (JVM), a lot of new languages that improve upon the developer experience with Java came up when development started: The closest contenders were Kotlin and Clojure (which is a Lisp dialect). We ultimately decided for Kotlin, as it is commercially supported by the company JetBrains, the developers of the popular IntelliJ IDE. Kotlin later became a first-class citizen on Android, contributing much to its popularity nowadays -- it is one of the fastest-growing languages currently on Github\footnote{See \url{https://octoverse.github.com/\#top-languages}.}. Kotlin offers a way more concise writing style than Java does, and introduces additional functional programming constructs. When comparing Kotlin and Clojure, Kotlin is notably more similar to Java, making it an appealing choice for this case due to the large number of Java programmers involved in the ImageJ community. In the JVM world, stable build systems and good dependency management have a long and successful history. In our case, we use Maven\footnote{See \url{https://maven.apache.org}} as a build system, which makes the build as easy as running the \texttt{mvn package} command, and is probably the most widespread build system used in the JVM ecosystem.

As mentioned before, our approach to developing the 3D Viewer replacement consists of two parts, \emph{scenery}, the rendering backend and framework, and \emph{sciview}, the user-facing frontend, which is the actual ImageJ/Fiji plugin. This split enables us to integrate new features into scenery, test them, and expose them later in sciview -- for example, support for distributed rendering (see Figure~\ref{fig:cave}) is implemented in scenery and usable, but not yet exposed in sciview. It also enables more stability for user-created scripts, where large backend changes in scenery are ``buffered'' by sciview, minimizing the number of changes that must be propagated to user scripts. While the user can choose to use pure scenery objects in her code, it is not necessary. We will continue with a description of the two parts, starting with scenery: 

\begin{figure}
    \centering
    \includegraphics[width=\linewidth]{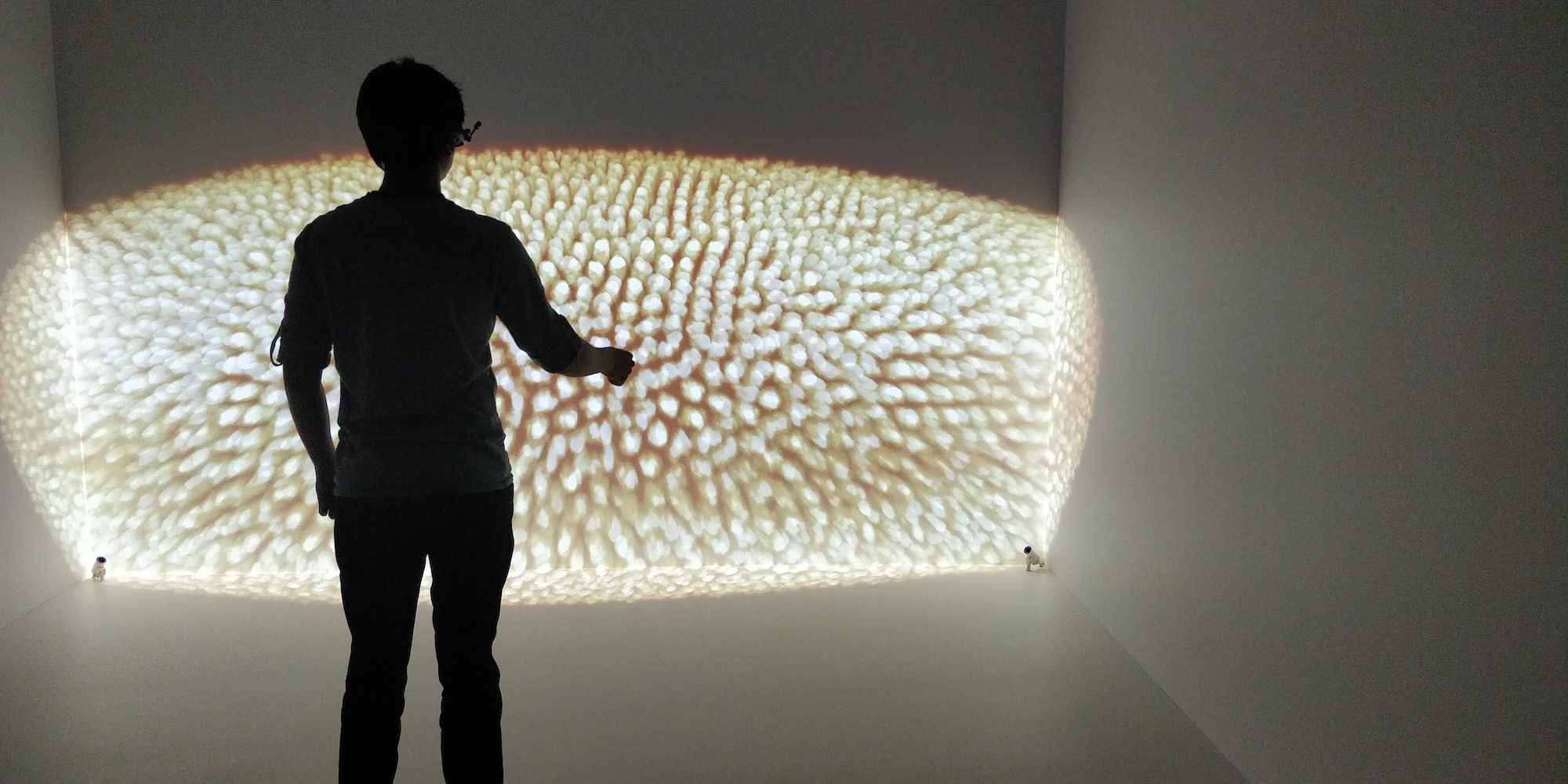}
    \caption{A volume rendering of a developing Drosophila embryo, rendered on a CAVE system. Distributed rendering support is handled by scenery, but not yet exposed in sciview. Image from \cite{Gunther:2019scenery}.\label{fig:cave}}
\end{figure}

In terms of rendering, scenery originally featured only an OpenGL 4.1-based renderer, with version 4.1 being chosen in order to support macOS. The Vulkan renderer was already being planned, and was added about a year after development started. The development of the Vulkan renderer led to a re-thinking of the renderer architecture, which led to improvements on both the Vulkan and the OpenGL side. In scenery, the renderers are completely decoupled from the rest of the library, such that the rendering-independent components can be tested on their own, even on systems without GPUs. The renderers are also interchangeable at runtime and feature a render graph architecture for the actual rendering pipeline, with the rendering pipelines also being exchangeable at runtime. To interface with native APIs, we use the LWJGL (Vulkan) and JOGL (OpenGL) libraries. While JOGL tries to wrap the OpenGL state machine into an object-oriented architecture, LWJGL provides access to the raw APIs. This means that in contrast to regular JVM applications,  memory management has to be handled by the developer and not by the garbage collector as usual. However, LWJGL provides excellent and fast APIs for memory management\footnote{See \url{https://blog.lwjgl.org/memory-management-in-lwjgl-3/}.}.

Scene contents are handled in a mostly traditional scene graph architecture, which we restricted to a scene tree \cite{boudier2015}, for more efficient parallel discovery of scenes with a large number of elements, as we require, e.g. for rendering large number of cells or neurons. 

Originally, scenery featured in-core volume rendering, being limited by the available memory of the GPU. In 2019, we have switched to an out-of-core volume rendering architecture based on technology developed from and in conjunction with the BigDataViewer \cite{pietzsch2015bigdataviewer} project. The technique implemented there is a combination the previously-published approach hierarchical blocking \cite{LaMar:1999Multiresolution,Beyer:2008Smooth} and the missing data scheme introduced in the BigDataViewer paper \cite{pietzsch2015bigdataviewer}. This enables us to not only render very large datasets, but also run filters on volumetric data on-the-fly.

%%% Key things:
% Maven and associated ecosystem is relevant here. Jitpack. Dependencies.
% Why to decouple scenery and sciview?
% Expose scenery objects to facilitate testing, but maintain a minimal sciview API such that scripts continue to work when major changes happen % in scenery, while still making the full api available. Additionally, imagej-mesh was introduced to decouple mesh data structures from scenery and sciview to allow for mesh-based processing in the absence of a graphical interface.

% [maybe cave figure here]

% ImageJ update sites

\section{Sustainability}

In this section, we want to detail decisions we have made in order to make the project sustainable in the longer run, split into technical aspects, and community/social aspects:

\subsection{The Technical Side: Kotlin, Polyglot Scripting, and Continuous Integration/Delivery}

With the choice of Kotlin as a language, we believe to have made a sustainable choice, for multiple reasons: First, should the JVM at some point not be sufficient anymore for our endeavours, Kotlin offers experimental support for compilation to native machine code\footnote{See \url{https://kotlinlang.org/docs/reference/native-overview.html}.} via LLVM \cite{Lattner:2004vw}. Second, Kotlin can also be compiled to JavaScript, enabling the possibility of a web-based viewer (which would require new rendering code, though) with the same basis. Third, all software written in Kotlin is completely interoperable with existing Java/JVM software, and to the developer it is mostly transparent whether Java or Kotlin was used. Kotlin libraries could even be decompiled to Java code.

In terms of extensibility, we offer a plugin-based system building on top of SciJava (\url{https://www.scijava.org}), and polyglot scripting of both scenery and sciview, with JavaScript, Jython, (J)Ruby, Matlab, BeanShell, Java, Scala, and Clojure being the currently available options. We hope this lowers the entry barrier to interacting with our software on a programmatic level (it will be interesting to see which language becomes the most-often used). The plugins required for interacting with all of these languages are developed in the SciJava community, are easily integrated, and do not need to be maintained by our team, which is a large plus.
Furthermore, with the rising popularity of Python, a Kotlin-Numpy bridge has also been developed by the community (\url{https://github.com/kotlin/kotlin-numpy/}), which we hope to be able to utilise in the future.

With the choice of Vulkan as an additional rendering API, we have stepped a bit into unknown territory, as it is a new API that has not been used extensively in scientific visualisation. Reasons for this choice were that Vulkan maps better to the current architecture of GPUs than OpenGL does, and provides a better developer experience, with excellent, and fine-grained debugging tools, like the Vulkan Validation Layers. These can be activated during development and perform runtime checks, which are mostly absent from the driver -- a stark contrast to OpenGL. Performance is enhanced in multiple ways: Vulkan offers much better multithreading support, and rendering/compute calls are send to the GPU by submitting work in batches via so-called command buffers, which can (and should!) be reused as much as possible. Finally, Vulkan offers more fine-grained synchronisation primitives than OpenGL. As with OpenGL, Nvidia offers interoperability with CUDA, e.g. for sharing buffers and images. 
The downside to using Vulkan is the up-front investment necessary: Vulkan differs significantly from OpenGL, is more verbose, and has a very steep learning curve -- especially synchronisation is something that bites rather often, and not all issues on that side can be caught reliably by the validation layers. All in all however, these downsides are more than balanced out by the improved development experience and  performance improvements. Programmers familiar with similarly explicit APIs like DirectX 12 or Metal will probably not find it difficult to learn. In the medium run, we plan on retiring the OpenGL renderer, as we see Vulkan as the more future-proof API. In the other direction, considering that Vulkan is already supported by currently 8 year-old Kepler-generation GPUs and later ones, we feel we also do not need to worry about backwards compatibility too much. We plan on adding a software renderer, probably using OSPray \cite{Wald:2017ee} in the future. Finally, Vulkan does not feature the strong coupling to the windowing system that OpenGL had anymore, so running on headless systems is much easier than with OpenGL.

For automated testing, we have created a Continuous Integration (CI) pipeline based on Travis CI (\url{https://travis-ci.org/}) that automatically runs unit tests and checks code coverage (via jacoco and \url{https://codecov.io}) on each commit to our git repositories. In addition, scenery includes a range of examples, which are also used as integration tests. At the moment, we are changing our CI pipeline to use Gitlab CI and automatically run these integration tests headlessly on different GPUs, and compare the resulting renderings with known-good images. Taking unit tests and integration tests together, we can reach a code coverage of about 70\%\footnote{On Github, around 25\% coverage is shown at the moment, because the Travis pipeline used at the moment does not yet run the tests requiring a GPU. Pending further testing, we will switch to the new CI pipeline in the near future.}. Testing code paths that require external hardware, such as HMDs, remain a problem, though. 

sciview (and with it, scenery) is delivered to the user through an ImageJ update site: Release versions are created manually, and in addition, we have a separate nightly update site, which is updated with each commit to the git repositories, facilitating Continuous Delivery (CD). The latter enables a fast response cycle for user-reported issues and for testing new developments with a limited set of users.

In the previous section, we have already mentioned the Maven-based build system -- in addition to providing an easy way to build our software, Maven also sets rather strict requirements for how the software is distributed: Maven-built software can be deployed to central repositories, such as Maven Central (\url{https://oss.sonatype.org}), with each released version of a software being immutable and digitally signed, contribution to both security and reproducability of builds. For versioning scenery and sciview, we use sementic versioning (\url{https://semver.org}), where major version changes indicate incompatible API changes, and minor version changes indicate backwards-compatible API changes.
Now, we cannot (and probably should not) create a new release for each commit to the repository, so in order to keep sciview up-to-date with respect to current developments in the scenery repository, we utilise JitPack (\url{https://jitpack.io}), where dependencies can be declared not only on release versions, but directly on Git commit hashes. The downside of this is that we cannot use semantic versioning for non-release builds, but we can guarantee that any build done in the past is going to work in the future, boosting reproducibility, even with versions still in development. The versions used, may it be the actual release version, or the Git commit used, are also shown to the user and in the log file, making error tracing easier for the developers.

% Kotlin! Kotlin-NumPy!
% CI, automated unit testing, automated integration testing, testing on GPUs, documentation, javadoc/kdoc
% CI: automatic deployment to update sites for both snapshots and releases,
% polyglot via scijava plugins

\subsection{The Social Side}

On the more soft side of things, we have found it slightly difficult to have people move away from 3D Viewer, even though both outdated and unmaintained, simply because they are used to it, and of course habits are difficult to change. A recent innovation in the ImageJ community, the image.sc forum for image analysis -- instead of the mailing list used over decades -- has alleviated that a bit, and led to more widespread knowledge about our project. We have also realised that the tool most often required is a simple piece of software users can just drop a single image or a time series onto. One of the authors was involved with the development of ClearVolume \cite{Royer:2015tg}, a visualisation tool originally intended to be used for live visualisation for microscopes, directly on the machine where the images are acquired. From the citations of the paper, we have noticed that ClearVolume is substantially more often used as Fiji plugin than as software for actual live visualisation, and that is probably because it makes it very simple to just view a single image or a time series.

\begin{figure*}
    \centering
    \includegraphics[width=\textwidth]{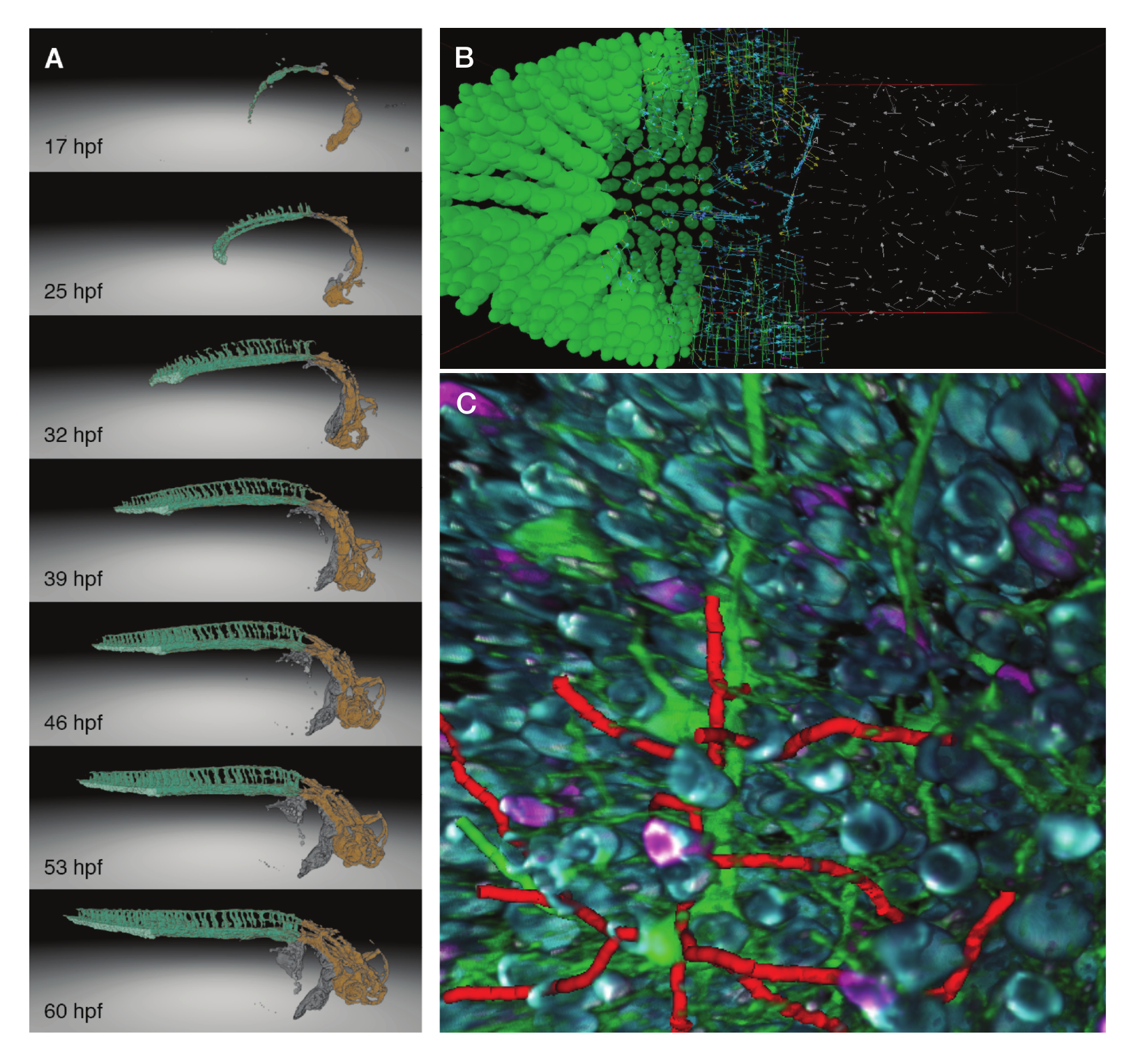}
    \caption{Example applications: A: Zebrafish vasculature timelapse, shown from 17 to 60 hours post fertilisation (hpf), using ambient occlusion to highlight smaller details; B: EmbryoSim, with forces acting on individual cells shown as errors, and cell bodies not shown towards the right; C/D: SNT for neuron tracing in volumetric images, with neurons shown in red, embedded in a dense volumetric image. See text for details.}
    \label{fig:examples}
\end{figure*}

In the meantime, according to GitHub, there are 18 projects using either scenery or sciview, with a six of them being our own. Exemplary usage of our tools include:

\begin{itemize}
	\item \emph{Multi-sample imaging and visualisation} \cite{Daetwyler:2019f45}, a paper using sciview to investigate, visualise, and analyse the development of vasculature in \emph{Danio rerio} (zebra fish) (see Figure~\ref{fig:examples}A),
	\item \emph{EmbryoSim}, a toolkit for the generation of plausible-looking \emph{Drosophila} (fruitfly) microscopy images for the training of machine learning algorithms for cell tracking and segmentation (see Figure~\ref{fig:examples}B),
	
	\item \emph{SNT}, a tool for neuron tracing, and successor to Simple Neurite Tracer \cite{Longair:2011snt}, which uses sciview as a viewer (see Figure~\ref{fig:examples}C), and
	\item \emph{Visualising regeneration in the mouse incisor}, a writeup in the popular biology blog, the Node, featured visualisations from sciview
	\footnote{See \sloppy\url{https://thenode.biologists.com/a-gnawing-question-which-cells-are-responsible-for-tooth-renewal-and-regeneration/research/}, and for the article \cite{Sharir:2019large}.}

\end{itemize}

In the latter two, the authors have been helping out with the development, an example of the tag team approach described in \cite{Reina:2020moving}.

For coordinated development, we hosted a hackathon dedicated to sciview development at the University of Idaho in 2018, and will again organise one in 2020, this time either at MPI-CBG in Dresden, or at CASUS in Görlitz. Hackathons have proven to be invaluable tools in the ImageJ community to drive development forward, and define new development goals. While we also heavily utilise online communication, e.g. via Gitter (\url{https://gitter.im/scenerygraphics/SciView/}), in-person meetings, with the associated socialising seem irreplaceable.

In order to keep development organised online, we utilise Github's Pull Request (PR) features heavily: All new features and bug fixes are submitted as a pull request before being merged into the master branch. This enables both code review and testing of code before something broken or inadequate reaches the main line code. For each PR, we run our CI tests on Windows, macOS, and Linux, and in addition run the code coverage and code quality tools. Issues in any PR are then resolved by the team members interactively, before the PR is finally merged, or rejected (with most PRs being actually accepted in the end).

% Community building, obstacles there
% The ImageJ community primarily interacts through: a mailing list, a forum shared with multiple image processing tools, dedicated chat rooms, and hackathons. Aside from the mailing list, we have adopted all of these mechanisms for supporting interaction within the sciview community, including hosting a dedicated sciview hackathon at the University of Idaho in 2019. 

% The development of plugins for ImageJ is not centrally controlled, and visualization tools have continued to be introduced into ImageJ during the development of sciview.

% Example of a biologist adopting sciview usage: Multisample SPIM of zebrafish vasculature [cite], example figure.

% attracting developers, SNT [figure], EmbryoSim [figure]

\section{Outlook}

While we believe our approach on the technical side is sound, there are still issues that we need to address:

\begin{itemize}
    \item \emph{Documentation} -- The JVM ecosystem forces the developer to document their code via Javadoc (Java) or Kdoc (Kotlin) by requiring dedicated documentation packages to be deployed to the central Maven package repository (see e.g. \url{https://javadoc.scijava.org/SciView}). We also require that new features in the form of Pull Requests on Github have adequate documentation. But as API-level documentation is not enough, we have started to write better, more explanatory documentation using Gitbook at \url{https://docs.scenery.graphics}. The documentation there is unfortunately far from complete. Already mentioned by \cite{Reina:2020moving}, software documentation often is the last part in the scientific process, and very unrewarding in the short-term, especially when the metric is "publications produced". However, there cannot be sustainable software development without adequate documentation.
    \item \emph{Funding} -- while the authors both develop and use scenery and sciview, and try to publish technology and papers based on them, it is sometimes difficult to acquire funding "just" for software, an issue already mentioned in \cite{Reina:2020moving}. Slowly, the large funding bodies, such as DFG in Germany, or the NIH in the US, are recognising the need for longer-term, stable software development, that sometimes also has to be done by the scientists requiring the software. We hope that this trend continues. Commercialisation might be an alternative option, but as we would ensure that our software stay open-source, it is not an easy task to derive a viable business model.
    \item \emph{Future prospects} -- The integration of sciview into the ImageJ ecosystem enables new technologies within the bioimage analysis community. By making the scenery framework accessible within ImageJ, the large number of existing ImageJ-based tools can now utilize virtual reality technologies. scenery's support for multiple rendering APIs introduces additional long-term stability as the landscape of computer graphics continues to evolve. New features introduced through sciview and scenery span from new image and mesh processing algorithms to out-of-core volume rendering, serving to extend beyond current ImageJ-based visualisation tools.
    \item \emph{Less Vis Gap} -- While we are mostly at home in the biomedical imaging community, we would like to interface better with the computer graphics community, to ensure that new algorithms and developments are incorporated into scenery and sciview.
\end{itemize}

\section{Conclusions}

In this paper, we have discussed the motivation and development process behind scenery and sciview, where sciview is intended to be the replacement for the current 3D Viewer in the ImageJ ecosystem. We have outlined how we believe language and ecosystem choice has an impact on sustainable development, and what tools we found valuable in supporting the developer. We have also described some major issues we are facing, that have already been discussed by \cite{Reina:2020moving}, and are common to a lot of visualisation software packages.

\section*{Acknowledgements}

We would like to thank all the contributors to the scenery and sciview projects, as well as Aryaman Gupta and Ivo F. Sbalzarini for proofreading this paper. We would also like to thank the Curtis Rueden for amazing help with development on the SciJava side, and the whole ImageJ community for their support so far. We further thank Simeon Ehrig and Tobias Huste from HZDR for their support in getting our headless CI tests going with different GPUs.

This work was partially funded by the Center of Advanced Systems Understanding (CASUS) which is financed by Germany’s Federal Ministry of Education and Research (BMBF) and by the Saxon Ministry for Science, Culture and Tourism (SMWK) with tax funds on the basis of the budget approved by the Saxon State Parliament.

\bibliographystyle{eg-alpha-doi}
\bibliography{TalesFromTheTrenches.bib}

\end{document}